# Title: Tuning Inelastic Light Scattering via Symmetry Control in 2D Magnet $CrI_3$

**Authors:** Bevin Huang[1†], John Cenker[1†], Xiaoou Zhang[2], Essance L. Ray[1], Tiancheng Song[1], Takashi Taniguchi[3], Kenji Watanabe[3], Michael A. McGuire[4], Di Xiao[2], Xiaodong Xu[1,5]*

[1]Department of Physics, University of Washington, Seattle, Washington 98195, USA
[2]Department of Physics, Carnegie Mellon University, Pittsburgh, Pennsylvania 15213, USA
[3]National Institute for Materials Science, 1-1 Namiki, Tsukuba 305-0044, Japan
[4]Materials Science and Technology Division, Oak Ridge National Laboratory, Oak Ridge, Tennessee 37831, USA
[5]Department of Materials Science and Engineering, University of Washington, Seattle, Washington 98195, USA

[†]These authors contributed equally to this work.

*Corresponding author's e-mail: xuxd@uw.edu

**Abstract: The coupling between spin and charge degrees of freedom in a crystal imparts strong optical signatures on scattered electromagnetic waves. This has led to magneto-optical effects with a host of applications, from the sensitive detection of local magnetic order to optical modulation and data storage technologies. Here, we demonstrate a new magneto-optical effect, namely, the tuning of inelastically scattered light through symmetry control in atomically thin chromium triiodide ($CrI_3$). In monolayers, we found an extraordinarily large magneto-optical Raman effect from an $A_{1g}$ phonon mode due to the emergence of ferromagnetic order. The linearly polarized, inelastically scattered light rotates by ~$40^o$, more than two orders of magnitude larger than the rotation from MOKE under the same experimental conditions. In $CrI_3$ bilayers, we show that the same $A_{1g}$ phonon mode becomes Davydov-split into two modes of opposite parity, exhibiting divergent selection rules that depend on inversion symmetry and the underlying magnetic order. By switching between the antiferromagnetic states and the fully spin-polarized states with applied magnetic and electric fields, we demonstrate the magnetoelectrical control over their selection rules. Our work underscores the unique opportunities provided by 2D magnets for controlling the combined time-reversal and inversion symmetries to manipulate Raman optical selection rules and for exploring emergent magneto-optical effects and spin-phonon coupled physics.**

**Main text:**

Raman scattering measures light inelastically scattered from collective quasiparticle excitations. Since it is highly sensitive to material parameters such as crystal symmetry and local electronic states, Raman spectroscopy has provided a powerful probe of a broad range of condensed matter phenomena, such as charge density waves[1], superconductivity[2], ferroelectricity[3], and topological physics[4]. In particular, Raman scattering from spin-phonon excitations has yielded incisive information on magnetic materials. For instance, in recently developed 2D van der Waals magnets, Raman scattering has been used to reveal magnetic order and phase transitions[5–7] down to a single layer[8–10].

Chromium triiodide ($CrI_3$), a van der Waals magnet, was shown to be a layered antiferromagnet in its few-layer form: spins within each layer are ferromagnetically (FM) coupled with strong out-of-plane anisotropy, while the interlayer exchange is antiferromagnetic (AFM)[11]. For bilayers, the system undergoes a spin-flip transition upon the application of a moderate magnetic field[11], switching from a layered AFM state to a fully spin-polarized state. In addition, magneto-optical effects manifest strongly and in distinctly novel ways in $CrI_3$. Examples include the very large magnetic-optical Kerr effect (MOKE)[11,12] and spontaneous helical light emission[13] from ferromagnetic monolayers, and electric-field induced Kerr rotation[14–16] and giant second-order nonreciprocal optical effects[17] in antiferromagnetic bilayers.

Furthermore, second harmonic generation measurements have revealed the restoration of inversion symmetry when the bilayer is switched from the AFM state to the fully spin-polarized state, highlighting the dependence of symmetry on the magnetic order in $CrI_3$ bilayers[17]. As such, Raman optical selection rules, and hence specific Raman modes, are likely to be controllable when switching between magnetic states. Considering the predicted and recently reported strong spin-lattice coupling in the monolayer[18,19], atomically thin $CrI_3$ is a promising candidate to study tunable magneto-optical Raman effects in the 2D limit.

In this work, atomically thin $CrI_3$ flakes were mechanically exfoliated onto oxidized silicon substrates and sandwiched between two flakes of hexagonal boron nitride to prevent their degradation in air. These samples were then cooled in a cold finger cryostat and excited at normal incidence using a HeNe laser at a wavelength of 632.8 nm, nearly resonant with the ligand-to-metal charge-transfer transition near 2 eV[13]. For Raman scattering measurements a power of 80 μW was used, while for polar reflectance magneto-circular dichroism (RMCD) measurements a power of 2 μW was used unless stated otherwise (see Methods for sample fabrication and measurement details).

We first present the Raman scattering results from monolayers of $CrI_3$. At 60 K, above the Curie temperature ($T_c$) of ~45 K, the monolayer is paramagnetic as confirmed by RMCD measurements shown in the inset of Fig. 1a. Raman spectra taken at this temperature in the co- and cross-linear polarization channels, which we denote XX and XY respectively, show four distinct peaks (Fig. 1a). Based on their observed polarization dependence as well as calculated phonon modes[18,20], we assign the peaks at 76.9 $cm^{-1}$ and 127.4 $cm^{-1}$ as scattering from $A_{1g}$ phonons and the peaks at 107.7 $cm^{-1}$ and 114.8 $cm^{-1}$ from two of the four $E_g$ phonons. The other two predicted $E_g$ phonons at around 50 $cm^{-1}$ and 230 $cm^{-1}$ are not seen, possibly due to their weak scattering cross-section.

When cooled below $T_c$, the monolayer becomes ferromagnetic, evident by the magnetic hysteresis in the RMCD measurements (Fig. 1b, left inset). As seen in Fig. 1b, there is a significant increase in the cross-linearly polarized Raman scattering from the 127.4 $cm^{-1}$ $A_{1g}$ phonon (see Extended Data Fig. 1 for data in the opposite spin orientation). The other $A_{1g}$ phonon at 76.9 $cm^{-1}$ and the two $E_g$ phonons show no marked change in their Raman signal in both the XX and the XY channels. As such, we will focus on only the 127.4 $cm^{-1}$ $A_{1g}$ phonon, corresponding to an out-of-plane and out-of-phase vibration between the two iodine layers (right inset of Fig. 1b), for the rest of our discussion on the $CrI_3$ monolayer.

To understand the sudden enhancement of the 127.4 $cm^{-1}$ peak in the XY channel, we track its intensity while changing the relative angle of the analyzing polarizer with respect to a fixed laser excitation polarization (green double-sided arrow along the 0º-180º axis, see Methods). Above $T_c$, the resultant polarization pattern at 60 K in Fig. 2a shows that this phonon mode exhibits $A_{1g}$ symmetry – the polarization axis, delineated by a dotted line, is co-linear with the laser excitation. In stark contrast, the polarization axis in the FM state at 15 K is rotated by some angle, $\varphi$, away from the excitation polarization (Fig. 2b). The rotation is roughly equal (about 40 degrees), opposite between the two time-reversal paired FM ground states and does not depend on the laser excitation polarization (Extended Data Fig. 2). In general, time-reversal symmetry breaking can induce a Hall-like, antisymmetric component in the Raman tensor of an $A_{1g}$ phonon (see Methods) that is responsible for the polarization rotation of the scattered light. But what is unexpected is that this rotation by a monolayer FM insulator is remarkably large, being two orders of magnitude larger than that from MOKE.

We also performed temperature-dependent measurements of the polarization pattern, starting at 60 K and cooling through $T_c$ without applying a magnetic field (Extended Data Figs. 3a-f). Figure 2c shows the extracted polarization rotation, $\varphi$, at select temperatures down to 15 K. $\varphi$ has an abrupt onset precisely when FM order is established at around 45 K and increases as the sample is further cooled down to 15 K. This temperature-dependent evolution of $\varphi$ matches the emergence of ferromagnetism seen in the remnant RMCD signal (Extended Data Fig. 3g). Combined with the fact that $\varphi$ was equal and opposite between the two FM states, we conclude that the origin of the polarization rotation is from the FM order in the monolayer.

Switching to the circular polarization basis further reveals the effects of FM order on the Raman scattering from the 127.4 $cm^{-1}$ $A_{1g}$ phonon. Figs. 2d-f show the helicity-resolved Raman scattering measurements from the 127.4 $cm^{-1}$ mode for the four possible scattering channels, $\sigma^+/\sigma^+$, $\sigma^+/\sigma^-$, $\sigma^-/\sigma^+$, and $\sigma^-/\sigma^-$, where the former describes the incident helicity and the latter describes the outgoing helicity ($\sigma^+/\sigma^+$ and $\sigma^-/\sigma^-$ scattering schematically drawn in the insets of Fig. 2d). In Fig. 2d, Raman scattering measurements above $T_c$ at 60 K show equal scattering in the $\sigma^+/\sigma^+$ and $\sigma^-/\sigma^-$ channels where the outgoing helicity is preserved. There is negligible scattering in the $\sigma^+/\sigma^-$ and $\sigma^-/\sigma^+$ channels where the outgoing helicity is reversed. In stark contrast, below $T_c$ at 15 K, Fig. 2e shows that with the magnetization pointing up, the 127.4 $cm^{-1}$ mode is dominated by the $\sigma^+/\sigma^+$ channel. By flipping the magnetization, the Raman scattering is then dominated by the $\sigma^-/\sigma^-$ channel (Fig. 2f), the exact time-reversed process of that observed when the magnetization pointed up.

Next, we explore the effects of magnetic order on the magneto-optical Raman scattering from bilayers of $CrI_3$. Unless stated otherwise, all the following measurements were performed at 15 K.

Unlike the monolayer, at zero magnetic field, two peaks distinct in energy appear at 126.7 $cm^{-1}$ and 128.8 $cm^{-1}$ in the XY and XX channels respectively (Fig. 3a). When the field is above the spin flip transition (0.7 T) to fully align the spins, only a single peak is observed at 128.8 $cm^{-1}$ in both the XX and XY channels (Fig. 3b). Figures 3c & e show the Raman intensity of these channels plotted as a function of magnetic field and Raman shift, while the magnetic field dependent RMCD signal is shown in Fig. 3d, providing information of the corresponding magnetic states.

Starting with the XX channel, the mode at 128.8 $cm^{-1}$ does not change in either energy or intensity as the magnetic field varies (Fig. 3e). In contrast, the XY channel in Fig. 3c shows that the 126.7 $cm^{-1}$ peak present in the AFM states is abruptly suppressed when the bilayer is switched to the fully spin-polarized states. Simultaneously, a peak at 128.8 $cm^{-1}$ emerges in the cross-polarized channel. The temperature-dependence of the peaks in the XY channel further confirms their magnetic origin. Above $T_N$ ~ 45 K, in the absence of an applied magnetic field, we only observe a single co-linearly polarized peak at 128.1 $cm^{-1}$ which slightly blueshifts to 128.8 $cm^{-1}$ as the bilayer is cooled to 15 K (Fig. 3g). Below $T_N$, both cross-polarized peaks at 126.7 $cm^{-1}$ (AFM state) and 128.8 $cm^{-1}$ (fully spin-polarized state) appear with the onset of magnetic order (Fig. 3f and Extended Data Fig. 4).

The appearance of the two Raman peaks in the AFM state and the dependence of the cross-linearly polarized spectra on the magnetic order can be understood by treating the $CrI_3$ bilayer as a coupled spring system[21,22]. In a monolayer, the $A_{1g}$ peak at 127.4 $cm^{-1}$ corresponds to a phonon mode describing an out-of-plane motion of the two iodine layers (Fig. 1b inset). In a bilayer, weak vdW interactions between the two layers split this mode into two (i.e. Davydov splitting). Since the lattice of bilayer $CrI_3$ is centrosymmetric, we can classify one of the two modes as an odd-parity mode ($u$), in which the layers vibrate out-of-phase as depicted in Fig. 3h, and the other as an even-parity mode ($g$) with in-phase vibrations between the layers (Fig. 3i). Since Raman scattering is an even-parity process, this implies that the even-parity $g$ mode is Raman-active, while the odd-parity $u$ mode is infrared-active but Raman-silent. This is consistent with the Raman spectra above $T_N$ (Fig. 3g).

Factoring in magnetic order leads to remarkable changes to the Raman optical selection rules. For the fully spin-polarized state, centrosymmetry remains intact (Fig. 3j), so it should behave exactly like the FM monolayers: the $g$ mode is active in both the XX and XY channels, while the $u$ mode remains silent. This is consistent with experimental findings. A comparison of the linear polarization patterns between the 127.4 $cm^{-1}$ $A_{1g}$ phonon mode in the FM monolayer and the $g$ mode in the bilayer shows virtually the same degree of polarization rotation (Extended Data Fig. 5). In the AFM state, parity no longer applies since the antiparallel spin configuration breaks inversion symmetry (Fig. 3k). Yet, the system remains invariant under the combined time-reversal and inversion symmetry, which forbids Raman activity in the Hall-like component in the Raman tensor of the $g$ mode. On the other hand, since the $u$ mode breaks inversion symmetry explicitly, it can be active in the XY channel, corresponding to the 126.7 $cm^{-1}$ peak.

To understand in detail the selection rules of these modes, we treat the bilayer $CrI_3$ as two weakly coupled FM monolayers. The Raman tensor of a monolayer can be decomposed into a spin-independent, diagonal part ($R_i$), and a spin-dependent, anti-symmetric part ($R_c$) (see Methods), with $R_i$ describing scattering in the XX channel and $R_c$ in the XY channel. If the magnetization is flipped, $R_i$ remains the same while $R_c$ changes sign. Under the weak coupling assumption, the Stokes and anti-Stokes component of the total induced electric dipole moment can be written as $p(t) = \sum_{l=1}^{2}(R_i^l + R_c^l)Q^l\, Ee^{i(\omega \pm \Omega)t}$, where $l$ is the layer index, $Q$ is the normal coordinate of the phonon, $E$ is the electric field, and $\omega$ and $\Omega$ are the photon and phonon frequencies, respectively.

In the AFM state, we have $R_i^1 = R_i^2 = R_i$ and $R_c^1 = -R_c^2 = R_c$. The normal coordinates are $Q^1 = -Q^2 = Q$ for the $u$ mode and $Q^1 = Q^2 = Q$ for the $g$ mode. The total induced dipole moment for the two modes are $p_{u,\text{AFM}} = 2R_cQ\, Ee^{i(\omega \pm \Omega)t}$ and $p_{g,AFM} = 2R_iQ\, Ee^{i(\omega \pm \Omega)t}$. Therefore, the *u* mode is active only in the XY channel, and the *g* mode in the XX channel. For the fully spin-polarized state, the key difference from the AFM state is that $R_c^1 = R_c^2 = R_c$. This leads to $p_{g,\text{FM}} = 2(R_i + R_c)Q\, Ee^{i(\omega \pm \Omega)t}$, and $p_{u,\text{FM}} = 0$. Consequently, the $g$ mode is active in both the XX and XY channels while the $u$ mode is silent. This analysis precisely matches our experimental observations. An analysis based on the magnetic point groups yields the same conclusion (see Methods).

Lastly, we exploit the control of the Davydov-split phonons by electrical switching of the magnetic states. Figures 4a & b show the schematic and optical microscope image of the bilayer device, respectively. The Raman intensity plot by sweeping the magnetic field near the spin-flip transition at an applied gate voltage $V_g$ of 0 V (5 V) is shown in Fig. 4c (d). The magnetic field at which the 126.7 $cm^{-1}$ peak is activated/suppressed is modulated from -0.7 T to -0.6 T, due to the electrical control of the critical magnetic field required to switch between the AFM and fully spin-polarized states[14,15]. Parking the magnetic field at -0.62 T, Figs. 4e & f compare the co- and cross-linearly polarized Raman spectra at two different $V_g$, 0 V and 5 V. The 126.7 $cm^{-1}$ peak is suppressed while the 128.8 $cm^{-1}$ peak is activated in the XY channel as a positive $V_g$ switches the magnetic states. In Fig. 4g, we continuously sweep the applied gate voltage from -1 V up to 6 V and monitor the Raman activity of the 126.7 $cm^{-1}$ and 128.8 $cm^{-1}$ phonons in the XY channel. The gradual suppression of the 126.7 $cm^{-1}$ phonon and the emergence of the 128.8 $cm^{-1}$ phonon track exactly the gate-dependent RMCD signal (Fig. 4h), with progressively larger negative RMCD signal as the bilayer is switched from an AFM state to the fully spin-down polarized state. From these measurements, we confirm the electrical switching of the 126.7 $cm^{-1}$ phonon mode through electrostatic control of the magnetic states, and thus the Raman optical selection rules, in a gated $CrI_3$ bilayer device.

In summary, we have observed a giant rotation (~$40^o$) of linearly polarized inelastically scattered light by a monolayer ferromagnetic insulator, established to originate from its magnetic order. While we have analyzed the optical selection rules based on symmetry, a quantitative understanding of the rotation calls for a microscopic theory that takes into account resonant Raman excitation. In bilayers of $CrI_3$, we demonstrated the Raman activation of a symmetry forbidden mode caused by the emergence of layered antiferromagnetic order. Coupling between the Raman selection rules and the combination of inversion symmetry and magnetic structure allows for the magnetoelectric switching of Davydov-split phonon modes. This leads to the activation/suppression of Raman activity for the odd-parity phonon mode and the magneto-optical rotation of scattered light from the even-parity phonon mode. These findings establish atomically thin $CrI_3$ as a unique platform for exploring externally sensitive magneto-optical effects through the exploitation of symmetries in the two-dimensional limit.

## Methods

### Sample preparation:

Bulk crystals of $CrI_3$ were mechanically exfoliated onto 285 nm $SiO_2$/Si substrates inside a glovebox with $N_2$ atmosphere. Monolayer and bilayer flakes were identified by optical contrast with respect to the substrate. Once suitable flakes were identified, they were encapsulated with 20-30 nm thick hexagonal boron nitride flakes.

Encapsulated $CrI_3$ samples were prepared inside the glovebox through a dry transfer technique using a poly(bisphenol A carbonate) (PC) film stretched over a polydimethylsiloxane (PDMS) cylinder as our stamp[23]. Each flake was picked up in this order and dropped onto a $SiO_2$/Si substrate: top hexagonal boron nitride (hBN), $CrI_3$ flake, followed lastly by bottom hBN.

Gated bilayer samples involved the addition of a 3-5 nm thick graphite flake so that the order of pickup went: hBN top dielectric, graphite contact, bilayer $CrI_3$, hBN bottom dielectric, and graphite bottom gate. The entire stack was then dropped onto 7-nm/70-nm thick V/Au contacts fabricated using a standard electron beam lithography technique.

**Optical measurements:**

All optical measurements were performed in a closed-cycle helium cryostat with a base temperature of 15 K utilizing the backscattering geometry. A superconducting solenoidal magnet was placed around the sample chamber such that magnetic fields of up to 7 T were applied in the Faraday geometry. An objective lens focused 632.8 nm light from a He-Ne laser down to a spot size of about 3 μm onto the sample at normal incidence.

For Raman scattering measurements, the scattered light was dispersed by a Princeton Acton 2500i spectrometer using a 1200 groove/mm diffraction grating and detected using a liquid nitrogen cooled charge-coupled device (CCD). Monolayer measurements used 80-μW of laser power and 5-minute integration times while 150 μW-and 3-minute integrations were used for the $CrI_3$ bilayers. BragGrate$^{TM}$ notch filters were used to reject Rayleigh scatter down to 5 $cm^{-1}$.

A linear polarizer and half-wave plate (HWP) placed directly after the notch filter allowed for collection polarization-dependent measurements. For our polarization-dependent measurements, we start in the co-linear (XX) polarization channel and rotate the HWP by 5° continuously until we scan through a full 360°. To save time, we only scanned through 180° in the temperature-dependent polarization dependence and duplicated this data for polarizations from 180° to 360°.

For magneto-Raman experiments, the applied magnetic field induces a Faraday rotation for all light that passes through optics within the magnet bore. By measuring the rotation of linearly polarized Rayleigh scattered light at various fields on $SiO_2$, we determined this Faraday rotation in our system to be ~5°/T. This rotation was compensated for with our polarization optics.

Magnetic materials may exhibit magnetic circular dichroism (MCD) which leads to a difference in the amplitude between reflected and transmitted right-circularly polarized (RCP) and left-circularly polarized (LCP) light. When an equal superposition of RCP and LCP, i.e. linearly polarized light, is incident on the material, the reflected and transmitted light will become elliptically polarized due to the MCD. In the reflection geometry, this effect is known as reflective magnetic circular dichroism (RMCD). RMCD measurements were performed using about 2 μW of power from a 632.8 nm He-Ne laser focused to a 3-μm beam spot. The experimental setup is similar to that used in previous RMCD measurements of the magnetic order in $CrI_3$[14].

**Symmetry analysis of the Raman tensor:**

The magnetic point group of monolayer $CrI_3$ with an out-of-plane magnetization is $D_{3d}(C_{3i}) = C_{3i} + \theta c_2' C_{3i}$, where $\theta$ is the time-reversal operator and $c_2'$ is an in-plane rotation axis. In the notation of Shubnikov and Belov, the magnetic point group is $\bar{3}m'$. To analyze the symmetry of the Raman tensor of the $A_{1g}$ mode, we break it into an $i$-tensor part ($R_i$) and a $c$-tensor part ($R_c$). The former is invariant under time-reversal, and thus is independent of the magnetic structure, while the latter changes sign upon time-reversal and is a consequence of the appearance of a macroscopic magnetization. Recalling that the Raman tensor must transform according to the same representation of the corresponding phonon mode, we find:

$$R = R_i + R_c = \begin{pmatrix} A & 0 & 0 \\ 0 & A & 0 \\ 0 & 0 & B \end{pmatrix} + \begin{pmatrix} 0 & C & 0 \\ -C & 0 & 0 \\ 0 & 0 & 0 \end{pmatrix} = \begin{pmatrix} A & C & 0 \\ -C & A & 0 \\ 0 & 0 & B \end{pmatrix}$$

$CrI_3$ bilayers are in the monoclinic phase. This is represented by the symmetry group, $C_{2h} = \{e, c_2', i, \sigma_\perp\}$, where $e$ is the identity operator, $c_2'$ is the two-fold rotation with an in-plane axis, $i$ is the inversion operator, and $\sigma_\perp$ is a reflection whose mirror plane is normal to $c_2'$.[17] The two Davydov-split phonon modes $g$ and $u$ transform under $C_{2h}$ as:

$$e|g\rangle = |g\rangle,\ c_2'|g\rangle = |g\rangle, \qquad i|g\rangle = |g\rangle,\ \sigma_\perp|g\rangle = |g\rangle$$

$$e|u\rangle = |u\rangle,\ c_2'|u\rangle = -|u\rangle, \qquad i|u\rangle = -|u\rangle,\ \sigma_\perp|u\rangle = |u\rangle$$

Applying the time-reversal operator has no effect on the phonon modes and leaves them unchanged. Considering magnetic structure, however, leads to distinct magnetic point groups between the fully spin-polarized and the AFM states. In the fully spin-polarized state, the magnetic point group is $C_{2h}(C_i) = \{e, i, \theta c_2', \theta\sigma_\perp\}$. The presence of inversion symmetry ($i$) renders the $u$ mode silent in both XX and XY channels and the lack of time-reversal ($\theta$) allows the Raman tensor for the $g$ mode to develop an antisymmetric component. For the AFM state, its magnetic point group is $C_{2h}(C_2) = \{e, c_2', \theta i, \theta\sigma_\perp\}$. Again by requiring that the Raman tensor transform according to the same representation of the corresponding phonon mode, we find that the $g$ mode is silent in the XY channel and the $u$ mode is silent in the XX channel.

**Acknowledgements:** This work was mainly supported by the Department of Energy, Basic Energy Sciences, Materials Sciences and Engineering Division (DE-SC0012509). Work at ORNL (MAM) was supported by the US Department of Energy, Office of Science, Basic Energy Sciences, Materials Sciences and Engineering Division. KW and TT acknowledge support from the Elemental Strategy Initiative conducted by the MEXT, Japan and JSPS KAKENHI Grant Numbers JP15K21722. BH acknowledges partial support from NW IMPACT. XX acknowledges the support from the State of Washington funded Clean Energy Institute and from the Boeing Distinguished Professorship in Physics.

**Author contributions:** XX, BH and JC conceived the experiment. BH and JC fabricated and characterized the samples, assisted by ELR and TS. BH and JC performed the Raman and magnetic circular dichroism measurements. BH, JC, XX, XZ, and DX analyzed and interpreted the results. TT and KW synthesized the hBN crystals. MAM synthesized and characterized the bulk $CrI_3$ crystals. BH, JC, ELR, DX, and XX wrote the paper with inputs from all authors. All authors discussed the results.

**Competing Interests:** The authors declare no competing financial interests.

**Data Availability:** The datasets generated during and/or analyzed during this study are available from the corresponding author upon reasonable request.

**Figures:**

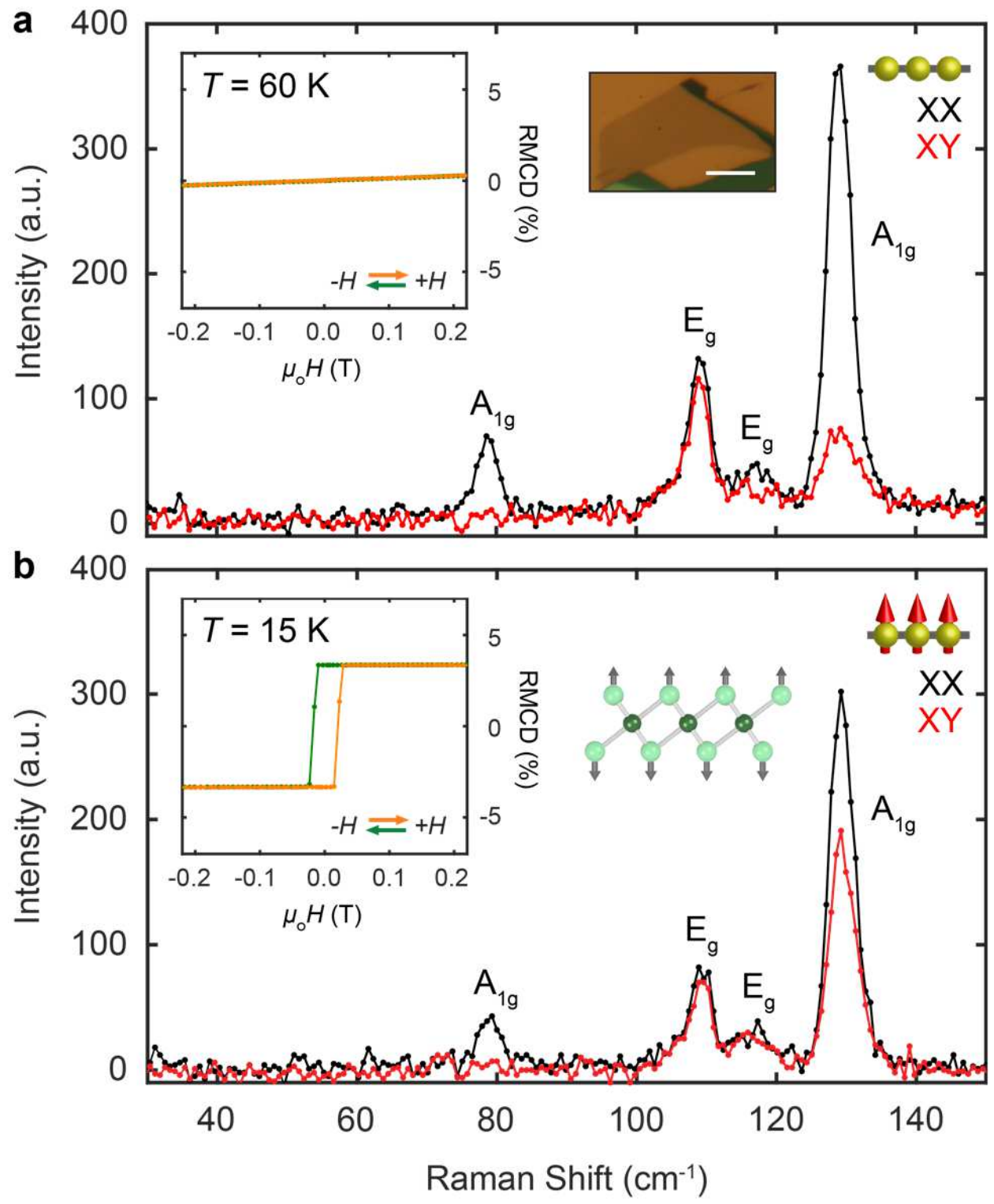

**Figure 1 | Raman scattering and its dependence on magnetic order in monolayer $CrI_3$. a-b,** Raman spectrum of monolayer $CrI_3$ with zero applied magnetic field in the **a,** paramagnetic state at 60 K and **b,** ferromagnetic state at 15 K. Black (red) curves correspond to co-linear (cross-linear) excitation and detection. Insets on the left show the RMCD signal as a function of magnetic field. Right inset of **a** is an optical micrograph of an exfoliated $CrI_3$ monolayer. Right inset of **b** shows the vibrational mode corresponding to the $A_{1g}$ mode at 127.4 cm$^{-1}$. Scale bar is 5 μm.

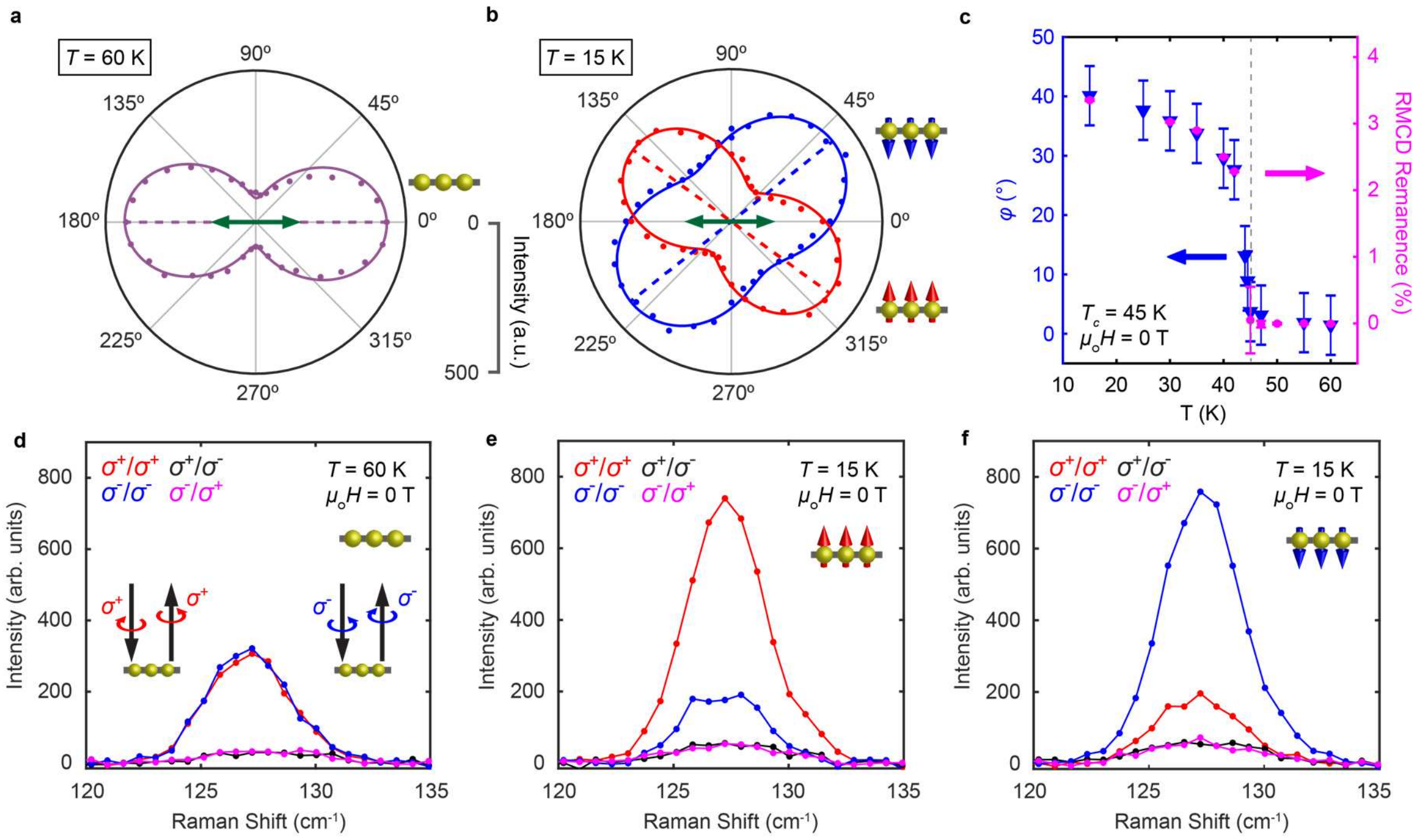

**Figure 2 | The effects of ferromagnetic order on the polarization of inelastically scattered light in monolayer $CrI_3$.** All data in this figure are taken at zero magnetic field. **a-b,** Polarization patterns of the 127.4 $cm^{-1}$ $A_{1g}$ mode in **a,** the paramagnetic state (purple) at 60 K and **b,** the two ferromagnetic states, spin-up (red) and spin-down (blue), at 15 K. The green arrow denotes the incident laser polarization. **c,** Raman scattering rotation, $\varphi$ (blue), and the RMCD remanence signal at zero applied magnetic field (magenta) plotted as a function of temperature. **d-f,** Helicity-resolved Raman spectra of the $A_{1g}$ peak in **d,** the paramagnetic state at 60 K and the two ferromagnetic states, **e,** spin-up and **f,** spin-down at 15 K. Insets in **d** depict the Raman scattering channels with identical helicities between the incident and scattered light.

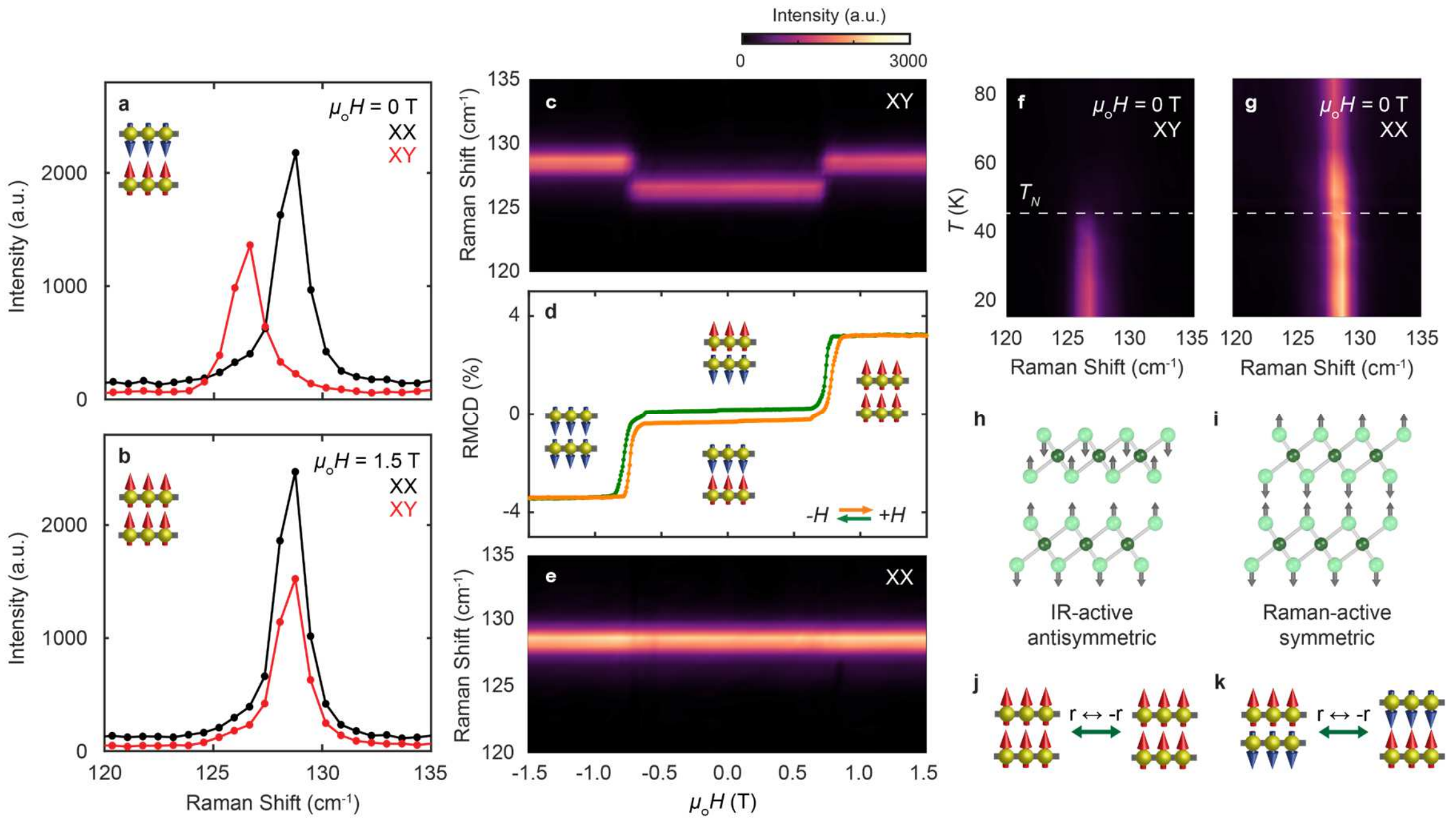

**Figure 3 | Coupling of magnetic order and Raman optical selection rules in bilayer $CrI_3$. a-b,** Co- (black) and cross-linearly (red) polarized Raman spectra taken in **a,** an AFM state at zero applied magnetic field and **b,** the fully spin-up polarized state at 1.5 T, with zero applied gate voltage. **c,** Color map of Raman spectra at a range of applied magnetic fields swept from 1.5 T to -1.5 T in the cross-linear scattering channel. **d,** Magnetic field dependent RMCD signal of the same bilayer. **e,** Color map of Raman spectra taken in the co-linear scattering channel**. f-g,** Color maps of **f,** cross-linearly polarized and **g,** co-linearly polarized Raman spectra taken at a range of temperatures while warming from 15 K to 85 K in zero applied magnetic field. The gray dashed line denotes the Neél temperature, $T_N$, where the mode in the cross-linear scattering channel is suppressed. From this, $T_N$ ~ 45 K. **i-j,** Illustrations of the two Davydov-split $A_{1g}$ modes in a $CrI_3$ bilayer: **h,** an IR-active peak at 126.7 cm$^{-1}$ and **i,** a Raman-active peak at 128.8 cm$^{-1}$. **j,** In the fully spin-polarized states, applying **-r** preserves the spin orientation, and therefore the centrosymmetry of the entire bilayer. **k,** On the other hand, applying the inversion operation, **-r,** in the layered AFM states switches the spin orientation of the two layers, thus breaking centrosymmetry. However, the combined time-reversal and inversion symmetry still holds.

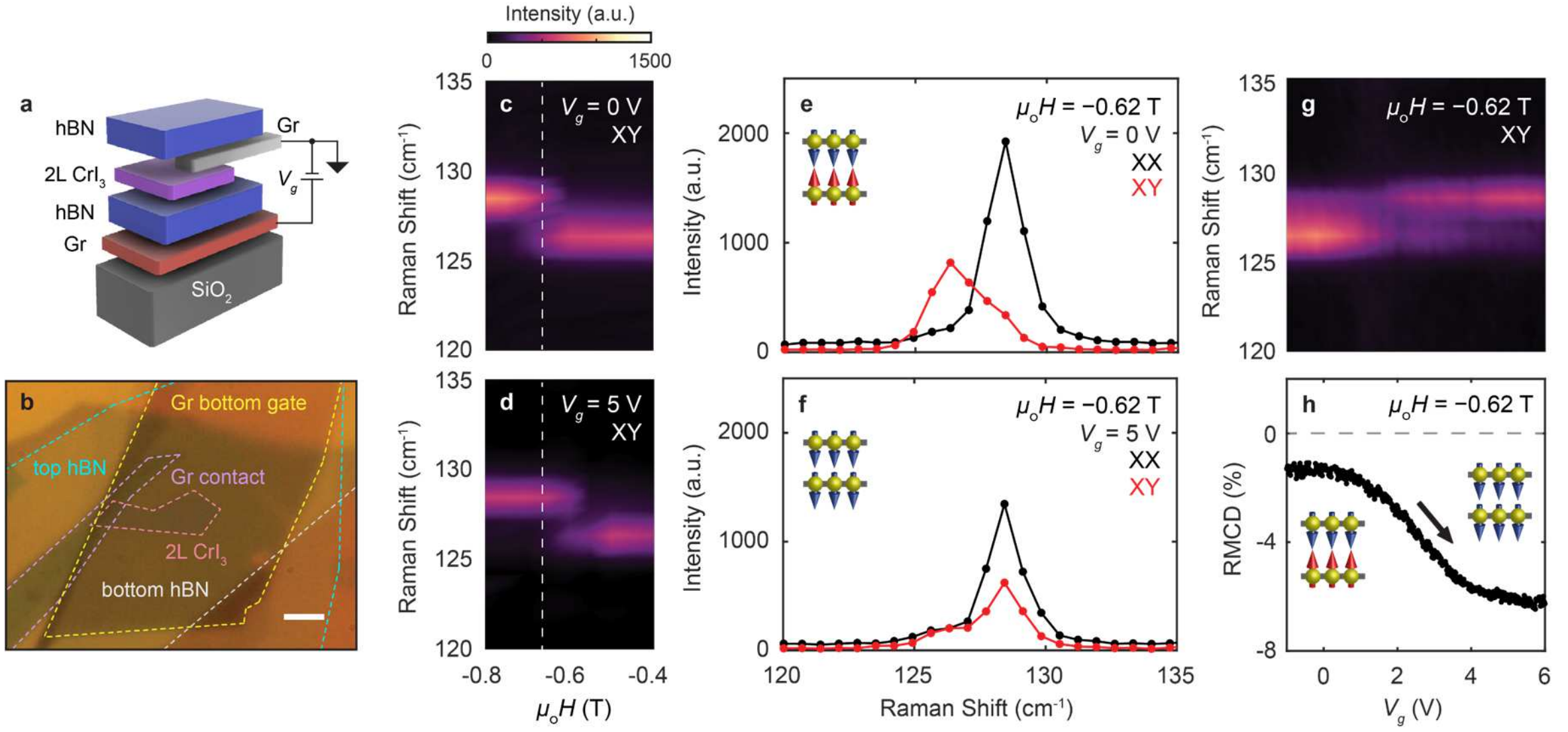

**Figure 4 | Electrical switching of a Raman-silent phonon in bilayer $CrI_3$. a,** Schematic of a gated $CrI_3$ bilayer device. **b,** Optical micrograph of the assembled gated $CrI_3$ bilayer device drawn in **a**. Scale bar is 5 μm. **c-d,** Color map of Raman spectra at a range of applied magnetic fields swept from -0.4 T to -0.8 T in the cross-linear scattering channel of a $CrI_3$ bilayer device taken at an applied gate voltage of **c,** 0 V and **d,** 5 V. The white dashed line denotes the magnetic field at which the spin-flip transition occurs in the ungated $CrI_3$ bilayer. **e-f,** Co-(black) and cross-linearly (red) polarized Raman spectra taken at an applied gate voltage of **e,** 0 V and **f,** 5 V. **g,** Color map of cross-linearly polarized Raman spectra taken at range of applied gate voltages from -1 V to 6 V at a fixed magnetic field of -0.62 T. **h,** RMCD signal of the same device taken in the same experimental conditions as in **g**.

# Supplementary information for

## Tuning Inelastic Light Scattering via Symmetry Control in 2D Magnet $CrI_3$

**Authors:** Bevin Huang[1†], John Cenker[1†], Xiaoou Zhang[2], Essance L. Ray[1], Tiancheng Song[1], Takashi Taniguchi[3], Kenji Watanabe[3], Michael A. McGuire[4], Di Xiao[2], Xiaodong Xu[1,5]*

[1]Department of Physics, University of Washington, Seattle, Washington 98195, USA
[2]Department of Physics, Carnegie Mellon University, Pittsburgh, Pennsylvania 15213, USA
[3]National Institute for Materials Science, 1-1 Namiki, Tsukuba 305-0044, Japan
[4]Materials Science and Technology Division, Oak Ridge National Laboratory, Oak Ridge, Tennessee 37831, USA
[5]Department of Materials Science and Engineering, University of Washington, Seattle, Washington 98195, USA

[†]These authors contributed equally to this work.

*Corresponding author's e-mail: xuxd@uw.edu

**Content:**

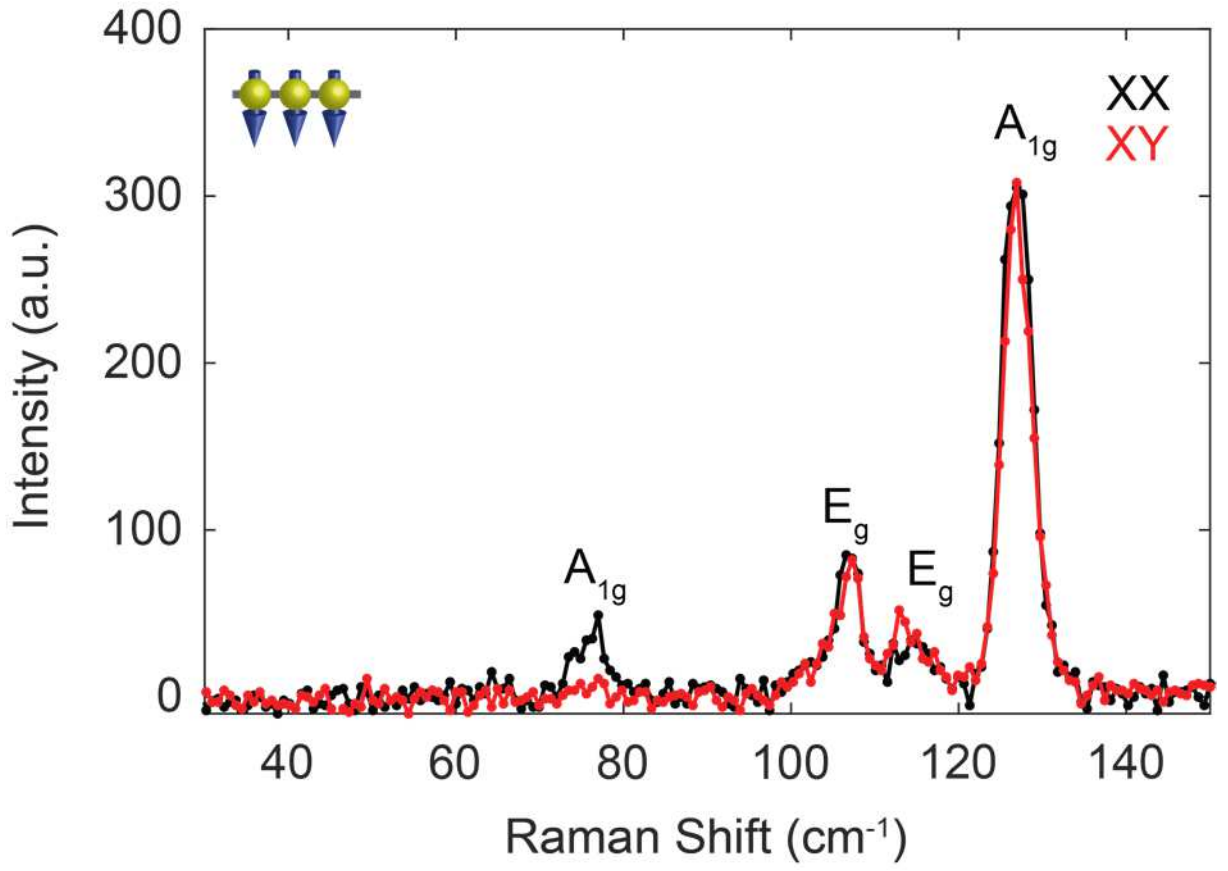

**Extended Data Fig. 1 | Polarization-resolved Raman scattering of monolayer $CrI_3$ in the spin-down ferromagnetic state.** Co- (black) and cross-linearly (red) polarized Raman spectra of a $CrI_3$ monolayer in the spin-down ferromagnetic state taken at 15 K with no applied magnetic field.

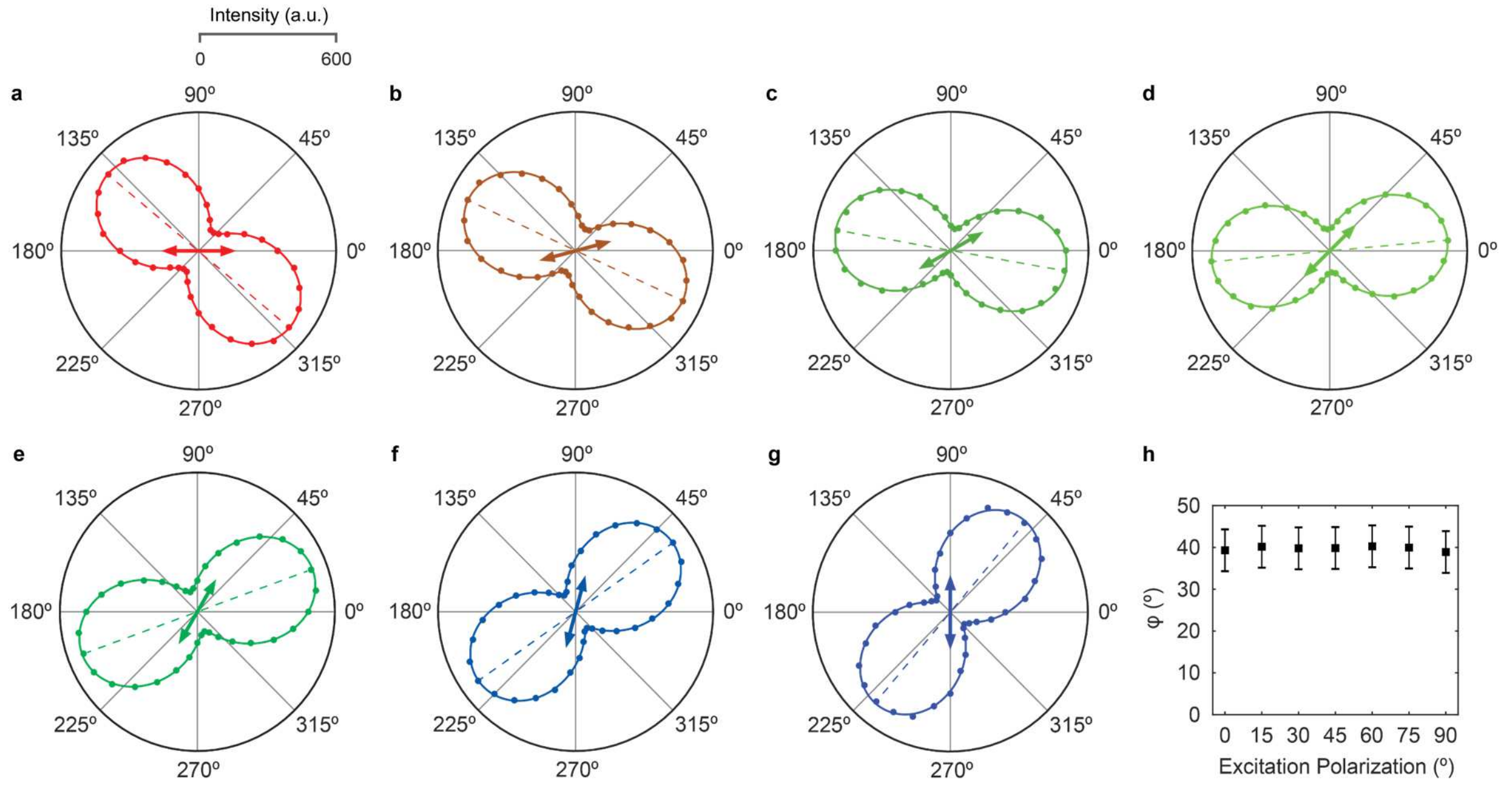

**Extended Data Fig. 2 | Independence of Raman scattering rotation on excitation linear polarization. a-g,** Polarization patterns of the 127.4 cm$^{-1}$ $A_{1g}$ mode in a $CrI_3$ monolayer for select incident laser polarizations denoted by the double-sided arrows. **h,** Raman scattering rotation, φ, of the 127.4 cm$^{-1}$ $A_{1g}$ mode in a $CrI_3$ monolayer as a function of the incident laser polarizations taken in **a-g**.

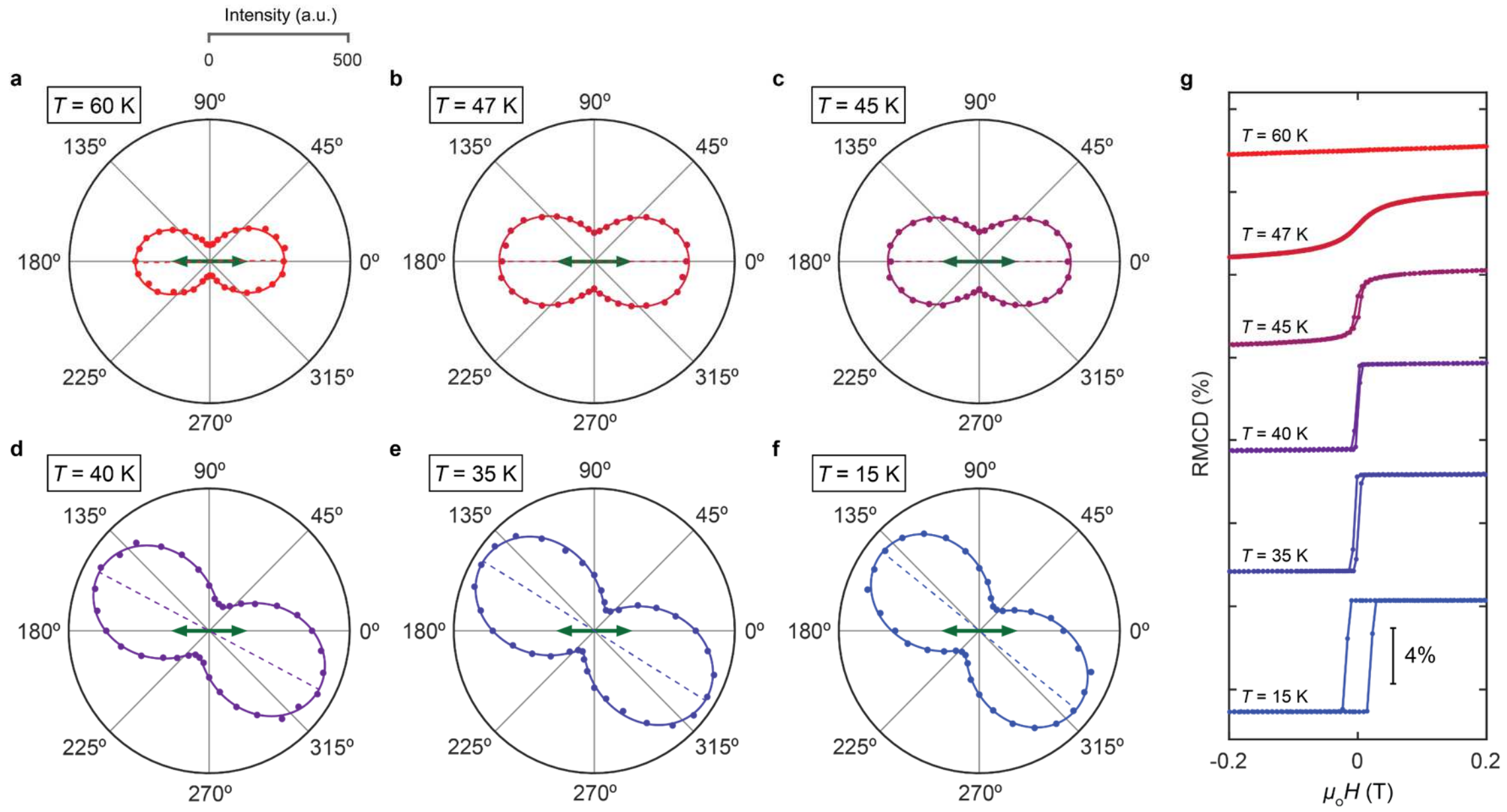

**Extended Data Fig. 3 | Temperature-dependent Raman scattering rotation and RMCD measurements. a-f,** Raman scattering polarization patterns of the 127.4 $cm^{-1}$ $A_{1g}$ phonon at select temperatures while cooling to 15 K from above $T_c$. Green double-sided arrows show the laser excitation polarization and dotted lines show the polarization axis of the resultant Raman scatter **g,** Magnetic field-dependent RMCD measurements at the same temperatures shown in **a-f.**

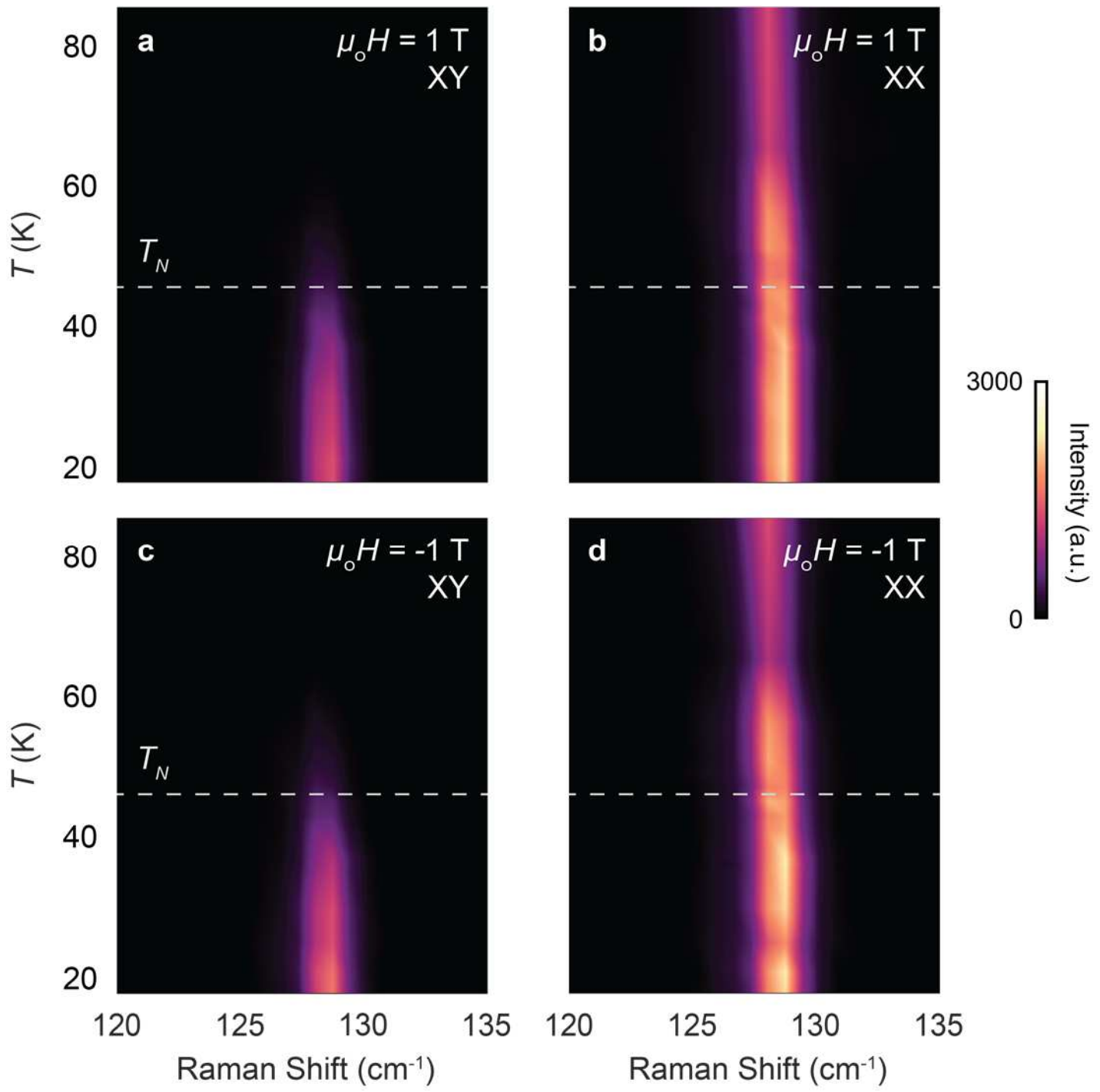

**Extended Data Fig. 4 | Temperature dependence of co- and cross-linearly polarized Raman scattered light from bilayer $CrI_3$ with an applied magnetic field. a-b,** Color maps of the temperature dependence of the **a,** cross- and **b,** co-linearly polarized Raman scattering channels with an applied magnetic field of 1 T. **c-d,** Identical measurements as in **a** and **b,** but in an applied field of -1 T for the **c,** cross- and **d,** co-linearly polarized Raman scattering channels.

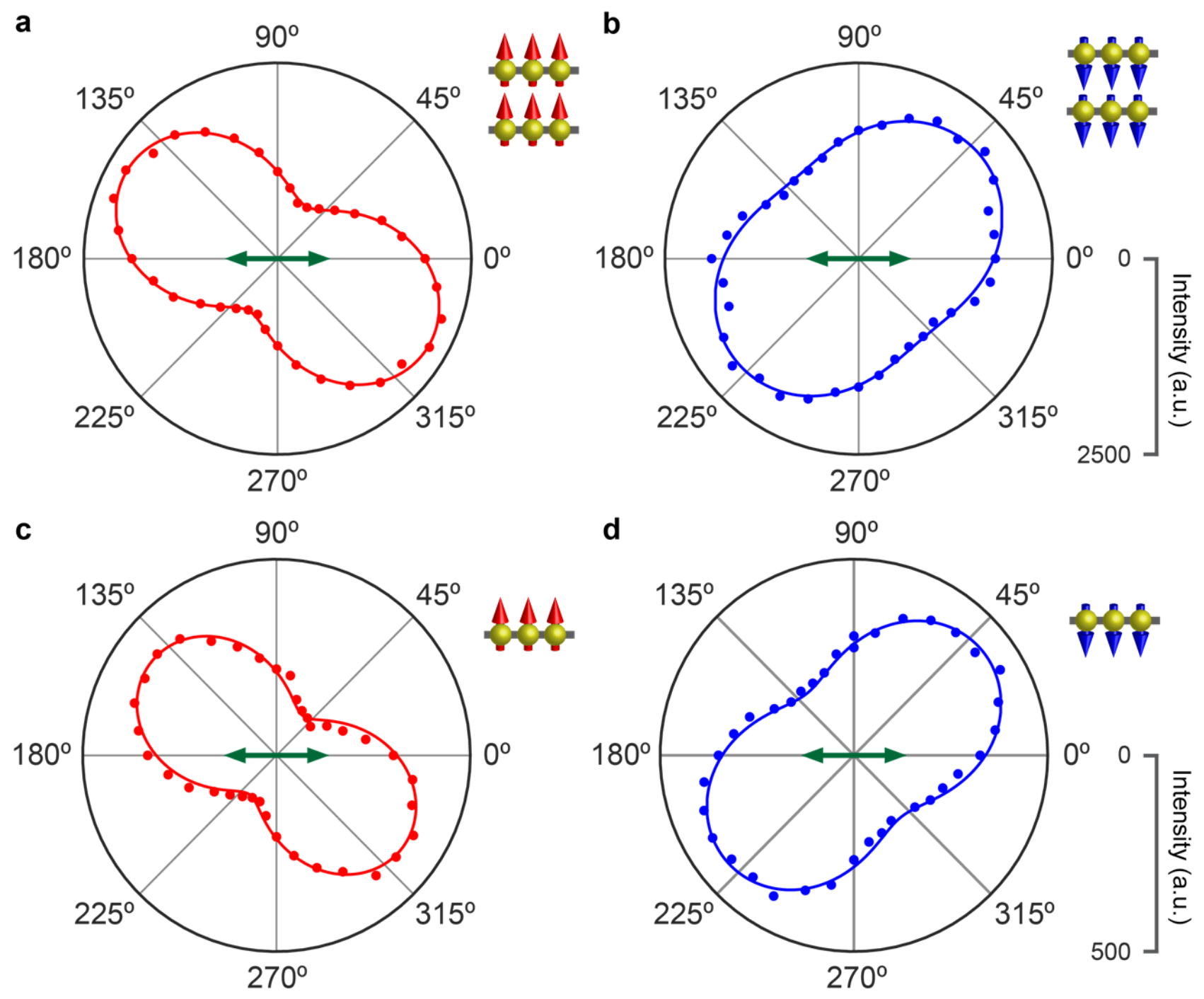

**Extended Data Fig. 5 | Raman scattering rotation from bilayer $CrI_3$ in the fully spin-polarized states. a-b,** Polarization patterns of the 128.8 $cm^{-1}$ mode in bilayer $CrI_3$ in the **a,** fully spin-up polarized and **b,** fully spin-down polarized states. **c-d,** Monolayer polarization patterns of the 127.4 $cm^{-1}$ phonon in **c,** the spin-up and **d,** spin-down states. The sign of the Raman scattering rotation in each spin state is the same and the magnitude of the rotation roughly equal between the monolayer and the bilayer. Green double-sided arrows show the laser excitation polarization.